\documentclass[a4paper]{article}

\usepackage{amssymb}
\usepackage{amsfonts}
\usepackage{amsmath}
\usepackage{graphicx}
\usepackage{hyperref}
\hypersetup{backref,
            pagebackref=true,
            ps2pdf,
            bookmarks=true,
            bookmarksnumbered=true,
            pdfauthor=McCabe Wydro,
            colorlinks=true,
            linkcolor=blue}

\title{Structure Constant of the Yang-Lee Edge Singularity}

\author{
\textrm{TOMASZ WYDRO}\\%
\textit{Laboratoire de Physique des Mat´eriaux}\\
\textit{ UMR CNRS 7556, Universit´e Henri Poincar´e} \\
\textit{Nancy 1, B.P. 239, F-54506 Vandoeuvre-les-Nancy Cedex, France}\\
wydro@lpm.u-nancy.fr
\\ \\
\textrm{JOHN F. McCABE}\\
\textit{412 Morris Ave., \# 34, Summit, NJ 07901,USA}\\
jfmccabe2@earthlink.net }

\begin{document}

 \maketitle

\begin{abstract}
This paper studies the Yang-Lee singularity of the 2-dimensional
Ising model on the cylinder via transfer matrix and finite-size
scaling techniques.  These techniques enable a measurement of the
2-point and 3-point correlations and a comparison of a measurement
of a corresponding universal amplitude with a prediction for the
amplitude from the $(A_{4}, A_{1})$ minimal conformal field theory.
\end{abstract}



\begin{center}
\bigskip
\end{center}

In 1985, Cardy \cite{IsingYLS} provided evidence that the Yang-Lee
edge singularity \cite{YL} of the 2-dimensional $(2D)$ Ising model
could be described by the non-unitary $(A_{4}, A_{1})$ minimal
conformal field theory (CFT) \cite{BPZ,Friedan} of the ADE
classification \cite{ADE}. Cardy's identification provided several
CFT predictions for the Yang-Lee edge singularity of the 2D Ising
model.  The predictions for the central charge, the exponent
$\nu$, and the low energy excitation spectrum have been confirmed
by numerical measurements on lattice spin models
\cite{Zuber-Uzelac-Mc}.

Cardy also determined the forms of 2-point and 3-point correlations
of the single primary field in the $(A_{4},A_{1})$ minimal CFT
\cite{IsingYLS}. These correlations define a universal amplitude,
which is known as a structure constant \cite{BPZ,Dotsenko}.  The
predictions of universal amplitudes are an important advance that
CFT brought to the understanding of critical points.  Since these
predictions of CFT are based on nontrivial applications of crossing
symmetries, they deserve to be experimentally tested \cite{BPZ}.

While lattice models have been used to numerically test some
predictions of universal amplitudes in unitary minimal CFTs \cite{
BPZ,Dotsenko,Barkema-vonGehlen}, no tests of such predictions have
been done for the non-unitary CFTs. This article tests such a
prediction for a non-unitary CFT by numerically measuring spin
correlations at the Yang-Lee edge singularity of the 2D Ising model.
The measurements enable a test of the CFT prediction of a universal
amplitude in the non-unitary $(A_{4}, A_{1})$ minimal CFT.

As pointed out by Cardy \cite{IsingYLS}, the non-unitary $(A_{4},
A_{1})$ minimal CFT has a single primary field $\phi(z,\bar z)$,
which has left and right conformal weights of -1/5 and a scaling
dimension $x$ of -2/5.  The 2-point and 3-point correlations of the
primary field $\phi(z,\bar z)$ have the respective forms:
\begin{equation}
G_{\phi\phi}(z_1, \bar{z_1}, z_2, \bar{z_2}) = |(z_1-z_2|^{4/5},
\label{2point}
\end{equation}and
\begin{equation}
G_{\phi\phi\phi}(z_1, \bar{z_1}, z_2, \bar{z_2}, z_3, \bar{z_3}) =
C|(z_1-z_2)(z_2-z_3)(z_3-z_1)|^{2/5}. \label{3point}
\end{equation}
With the Coulomb gas formalism \cite{Dotsenko}, Cardy showed that
the single structure constant, $C$, of the non-unitary $(A_{4},
A_{1})$ minimal CFT has the value \cite{IsingYLS}:
\begin{equation}
C=\sqrt{-\frac{[\Gamma(6/5)]^2\Gamma(1/5)\Gamma(2/5)}{\Gamma(3/5)[\Gamma(4/5)]^3}}
\label{eq:StructConstExact}
\end{equation}
In this article, numerical measurements at the Yang-Lee edge
singularity check this CFT prediction.

The numerical measurements were made for the 2D ferromagnetic Ising
model whose Hamiltonian, $H$, is:
\begin{equation}
H=-\sum_{j=1}^{M}\sum_{i=1}^{N}[J(S_{i,j}S_{i,j+1}+S_{i,j}S_{i+1,j})+hS_{i,j}]
. \label{eq:IsingHamilt}
\end{equation}
In the ferromagnetic Hamiltonian, the spin-spin coupling $J$ is
positive.  For the Ising model, the Yang-Lee edge singularity
occurs at temperatures above the critical temperature and for
purely imaginary values of the magnetic field, $h$, i.e., $h = iB$
with $B$ real \cite{YL}. In particular, spin correlations were
measured at a temperature, T, for which $J/k_BT = 0.1$

For the above Hamiltonian, the transfer matrix was used to
evaluate the correlations. In particular, 2-spin and 3-spin
correlations were evaluated on torii of length, $M$, and of
various diameters, $N$. In these evaluations, $M$ was always much
larger than $N$, i.e., M = 512 and N = 3 - 8.  Thus, the measured
correlations had the same distance behavior as correlations on an
infinitely long cylinder when distances between spin fields were
very small compared to $M$.

From the numerical evaluations, finite-size scaling provided the
tool for extracting values of physical properties in the
thermodynamic limit \cite{FiniteScall}. In particular, the spin
correlations are measured at special values of the purely
imaginary magnetic field, $h(N) = iB_{YL}(N)$.  Each special value
, $B_{YL}(N)$, satisfies the phenomenological renormalization
group (PRG) equation for infinite cylinders of diameters $(N-1)$
and $N$:
\begin{equation}
\frac{\xi(iB_{YL}(N), N-1)}{N-1}=\frac{ \xi(iB_{YL}(N), N)}{N}\text{
.} \label{PRG}
\end{equation}
In the PRG equation, $\xi(iB, N)$ is the spin-spin correlation
length on the infinite cylinder of \mbox{diameter} $N$ when the
magnetic field is $iB$. The PRG equation imposes that the
spin-spin correlation length scales linearly with $N$ as
$N\to\infty$. When evaluated at PRG values of the magnetic field,
other physical quantities will scale with the width, $N$, to their
values in the thermodynamic limit, i.e., near the Yang-Lee edge
singularity \cite{Derrida,Zuber-Uzelac-Mc}.

On an infinite cylinder of width $N$, CFT predicts that
correlations will depend exponentially on the distances between
the fields in the correlations when said distances are large
compared the cylinder's diameter, $N$ \cite{Cardy2}. For a 2-point
correlation, the correlation length is related to the scaling
dimension, $x$, of the two conformal fields of the correlation. In
particular, the 2-point correlation has the form $\exp(-2\pi x
(y_1 - y_2)/N)$ when $|y_1 - y_2| >> N$. Here, $y_1$ and $y_2$ are
the positions of the two fields of the correlation along the axis
of the infinite cylinder. For the 3-point correlation, the
exponential behavior on the distances between the fields of the
correlation is again fixed by the scaling dimensions, $x's$, of
the various fields of the correlation.

At the Yang-Lee edge singularity of the 2D Ising model, we used the
amplitudes of the 2-spin and 3-spin correlations, i.e., $A_{ss}$ and
$A_{sss}$, respectively, to evaluate the 3-spin structure constant.
In particular, the measured values of the 3-spin structure constant,
$C(N)$, were obtained from the relation:
\begin{equation}
C(N)=\frac{A_{sss}(iB_{YL}(N))}{[A_{ss}(iB_{YL}(N)]^{3/2}}.
\end{equation}
Here, $A_{ss}(iB_{YL}(N))$ and $A_{sss}(iB_{YL}(N))$, are the
measured amplitudes of the 2-spin and 3-spin correlations,
respectively, at PRG values of the magnetic field. Similarly, we
used the PRG measurements of the correlation length $\xi(N)$ to
measure the conformal dimension, $x$, of the spin field, i.e.,
$x(N) = N/[2\pi\xi(iB_{YL}(N))]$. The scaling behavior of these
physical quantities with the cylinder's width, $N$, was used to
obtain their values in thermodynamic limit, i.e., as $N\to\infty$.

Table \ref{tab:CriticalFields} summarizes our transfer matrix
results\footnote{While the $B_{YL}(N)$'s for N = 3 - 7 were
evaluated from the PRG equation, $B_{YL}(N)$ for N = 8 was only
approximately evaluated for numerical reasons. $B_{YL}(8)$ was
determined from the $B_{YL}(N)$'s for N = 3 - 7 by assuming a
leading finite-size scaling behavior.} for $M = 512$ and $J/k_BT = 0.1$.
\begin{table}[h]
\vspace{1mm} \centering
\begin{tabular} {clll}
\hline
 $N$ &  $B_{YL}(N)$ & $x(N)$ & $|C(N)|$ \\
\hline
3 & 0.184802 & 0.353929 & 1.80838\\
4 & 0.183348 & 0.376870 & 1.83711\\
5 & 0.183064 & 0.385748 & 1.85736\\
6 & 0.182982 & 0.390108 & 1.87054\\
7 & 0.182951 & 0.392693 & 1.87937\\
8 & 0.182946 & 0.392911 & 1.88633\\
\ldots&\ldots&\ldots&\ldots\\
$\infty$&\ldots&0.398(2)&1.923(13)\\
$CFT$& --- &0.4&1.9113 \\
\hline
\end{tabular}
\caption{PRG Measurements of conformal dimension and structure
constant. \label{tab:CriticalFields}}
\end{table}

In these measurements of correlations, adjacent spin fields were
separated by distances that were large compared to $N$ and small
compared to $M$.  In particular, we measured correlations in which
the distances between adjacent spin fields were between about $4N$
and $0.06M$. For such distances, the CFT prediction of an
exponential dependence on the distances between fields was obtained.
In Table \ref{tab:CriticalFields}, the $\infty$ line shows values
obtained by extrapolating our measurements to the thermodynamic
limit, i.e., $N = \infty$. The extrapolated values were obtained
from nonlinear fits of the measured values of $x(N)$ and $|C(N)|$ to
functions of form $f(N) = f(\infty)+ f_1 N^{-\alpha} $, i.e., to
account for leading finite-size corrections. In each such
extrapolation, the coefficient, $f_1$, and the scaling power
$\alpha$ were determined by finding the best fit to the measured
data points. The best fits were found when the scaling corrections
or $x(N)$ and $|C(N)|$ had exponents, $\alpha$, of about 2.4 and
1.2.
In Table \ref{tab:CriticalFields}, the last line shows Cardy's CFT
predictions for $x$ and $|C|$ from the $(A_{4}, A_{1})$ non-unitary
minimal CFT as the model.

\begin{figure} [h]
 {\centering\includegraphics[scale=1.2]{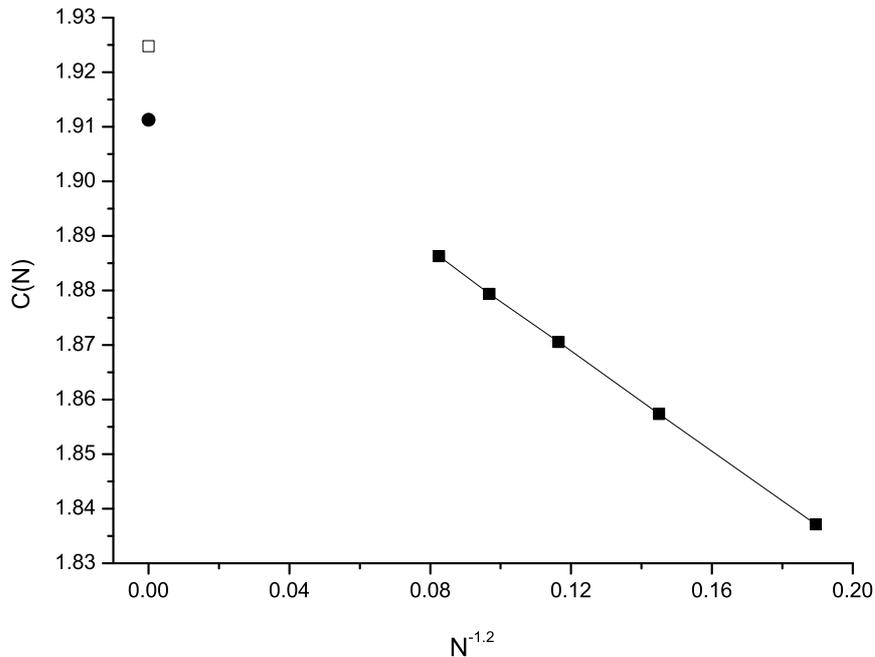}}
\caption{The measured structure constant (squares) for cylinders
of diameter $N$ plotted against nonlinear fit for $N= 4 - 8.$
\label{fig:StructConstScaling}}
\end{figure}

Figure \ref{fig:StructConstScaling} plots our  data for $|C(N)|$
and our best fit (line) based on a leading correction in
$N^{-1.2}$. In Figure 1, the black squares represent the measured
$C(N)$'s, the empty square represents the value of $C(\infty)$
obtained from the best fit to the measurements, and the black
circle represents Cardy's CFT prediction for $C(\infty)$.

The measured structure constants for cylinders of diameter 3 - 8
exhibits an approximate finite-size scaling correction of the form
$N^{-1.2}$.  The finite-size scaling motivated fit matches the
measured values of structure constant well except for the thinnest
cylinder, $N=3$.  There, one is not surprised that a large $N$
expansion does not work well.  Finally, the dot shows the
structure constant predicted by CFT.

In conclusion, our finite-size scaling measurements for the 2D Ising
model at the Yang-Lee edge singularity produce a value of the
structure constant that agrees well with that of the $(A_4, A_1)$
minimal CFT.  These measurement confirm the CFT prediction of this
universal amplitude at the Yang-Lee edge singularity of the 2D Ising
model.


\bigskip



\begin{thebibliography}{99}

\bibitem{IsingYLS} J.L. Cardy, Phys. Rev. Lett. {\bf 54}, 1354 (1985).

\bibitem{YL} C.N. Yang and T.D. Lee, Phys. Rev. {\bf 87}, 404 and 410 (1952).

\bibitem{BPZ} A.A. Belavin, A.M. Polyakov, and A.B. Zamolodchikov, Nucl.
Phys. {\bf B241}, 333 (1984).

\bibitem{Friedan} D. Friedan, Z. Qiu, and S. Shenker, Phys. Rev. Lett. {\bf 52}, 1575
(1984), and Comm. Math. Phys. {\bf 107}, 535 (1986).

\bibitem{ADE} A. Cappelli, C. Itzykson, and J.-B. Zuber, Nucl. Phys. {\bf B280},
445 (1987), and Comm. Math. Phys. {\bf 113}, 1 (1987); A. Kato,
Mod. Phys. Lett. {\bf A2}, 585 (1987); for a review see C.
Itzykson and J.-M. Drouffe, {\it Statistical Field Theory}
(Cambridge University Press, U.K., 1989) Ch. IX.

\bibitem{Zuber-Uzelac-Mc} C. Itzykson, H. Saleur, and J.-B. Zuber, Europhys. Lett.
{\bf 2}, 91 (1986); K. Uzelac and R. Jullien, J. Phys. {\bf A14},
L151 (1981); J.F. McCabe and T. Wydro, Int. J. of Mod. Phys. {\bf
B20}, 495 (2006).

\bibitem{Dotsenko} Vl.S. Dotsenko and V.A. Fateev, Nucl. Phys. {\bf B240}, 312 (1984),
and {\bf B251}, 691 (1985).

\bibitem{Barkema-vonGehlen} G.T. Barkema and J. McCabe, J. Stat.
Phys. {\bf 84}, 1067 (1996); J. McCabe and T. Wydro, Int. J. of Mod.
Phys. {\bf A13}, 1013 (1998); G. von Gehlen, V. Rittenberg, and T.
Vescan, J. Phys. {\bf A20}, 2577 (1987).

\bibitem{FiniteScall} M.E. Fisher and M.N. Barber, Phys. Rev. Lett.
{\bf 28}, 1516 (1972); for a review see M. Henkel, {\it Conformal
Invariance and Critical Phenomena} (Springer-Verlag, Germany,
1999) Ch. 3.

\bibitem{Derrida} M.P. Nightingale, Physica {\bf A83}, 561 (1976); and
in {\it Finite Size Scaling and Numerical Simulation of
Statistical Systems}, edited by Ed. V. Privman (World Scientific,
Singapore, 1990) Ch. VII.

\bibitem{Cardy2} J.L. Cardy, J. Phys. {\bf A17}, L385 (1984), and
Nucl. Phys. {\bf B270}, 186 (1986); for a review see M. Henkel,
{\it Conformal Invariance and Critical Phenomena}
(Springer-Verlag, Germany, 1999) Ch. 13.


%
%
%
%

\end{thebibliography}
\end{document}